\documentclass[conference,a4]{IEEEtran}
\usepackage{graphicx,graphics}
\usepackage{wasysym}
\usepackage{csquotes}
\usepackage{cases}
\usepackage{cite}
\usepackage{array}
\usepackage{siunitx}
\newcommand{\DeltaE}{\ensuremath{\Delta E}}
\newcommand{\kB}{\ensuremath{k_{\mathrm{B}}}}
\newcommand{\Vc}{\ensuremath{V_{\mathrm{c}}}}
\newcommand{\Vpulse}{\ensuremath{V_{\mathrm{pulse}}}}
\newcommand{\Tpulse}{\ensuremath{T_{\mathrm{pulse}}}}
\newcommand{\BER}{\ensuremath{{\mathrm{BER}}}}
\newcommand{\BERLSBs}{\ensuremath{\mathrm{BER}_{\mathrm{LSBs}}}}
\newcommand{\BERHSBs}{\ensuremath{\mathrm{BER}_{\mathrm{HSBs}}}}
\newcommand{\TMR}{\ensuremath{{\mathrm{TMR}}}}

\begin{document}

\title{Use of Magnetoresistive Random-Access \\ Memory as  Approximate Memory \\ for  Training  Neural Networks   
}

\author{
	\IEEEauthorblockN{ Nicolas Locatelli, Adrien F. Vincent ̃and  Damien Querlioz}
	\IEEEauthorblockA{Centre for Nanoscience and Nanotechnology, CNRS, Univ. Paris-Sud, 91405 Orsay, France.\\
					  Email: damien.querlioz@u-psud.fr}
}


\maketitle

\begin{abstract}
Hardware neural networks that implement synaptic weights with embedded non-volatile memory, such as spin torque memory (ST-MRAM), are a major lead for low energy artificial intelligence. In this work, we propose an approximate storage approach for their memory. We show that this strategy grants effective control of the bit error rate by modulating the programming pulse amplitude or duration. Accounting for the devices variability issue, we evaluate energy savings, and show how they translate when training a hardware neural network. On an image recognition example, $74\%$ of programming energy can be saved by losing only $1\%$ on the recognition performance.
\end{abstract}


\section{Introduction}

The development of systems implementing machine learning applications, typically relying on artificial neural networks, has become a major goal of microelectronics research \cite{ dean2018new,yu2018neuro}.
Interestingly, such systems, which require high amounts of memory, demonstrate relative resilience to inaccuracies during computation \cite{burr2015experimental}.
Data integrity in memories is ensured with significant costs in energy and performance of VLSI systems~\cite{sampson_good-enough_2013, palem_ten_2013}, and these costs are exacerbated as memory devices move to smaller dimensions, along with increasing process variations\cite{indaco_impact_2014, worledge_switching_2010}. 
Therefore, the approximate computing paradigm~\cite{han_approximate_2013, kim_inexact_2011, kirsch_incorrect_2012, sampson_good-enough_2013, venkataramani_approximate_2015}, which  proposes to trade off computational accuracy for higher energy efficiency~\cite{chippa_scalable_2014}, could be a lead for reducing the energy costs related to memory in hardware artificial neural networks.  

This idea takes  particular meaning as many hardware artificial neural networks propose using emerging types of memories for implementing synaptic weights \cite{yu2018neuro}.
In particular, with considerable progress made in recent years, spin  torque magnetic random access memories (ST-MRAMs) are highly attractive, as they provide fast programming and outstanding endurance in a compact and non-volatile memory cell, fully embeddable at the core of CMOS ~\cite{locatelli_spintronic_2015}.

In this work, we explore the unique properties of ST-MRAMs in the framework of approximate computing, and propose how they could be exploited in  hardware neural networks.
Relying on recently developed analytical models of the ST-MRAMs programming statistics, this work first explores the relation between programming precision and energy expenditure.
We  show using circuit simulations that the consequences of device variability increase dramatically as the precision requirements become stronger. 
We then investigate how allowing higher error rates can provide significant energy savings, 
and show based on an example that this approach suits judiciously to data storage in the context of artificial neural network applications.

\section{Stochastic Programming of the Magnetic Bit}
\label{Sec:programming}

\begin{figure*}[ht]
	\centering
	\includegraphics[width=0.9 \linewidth]{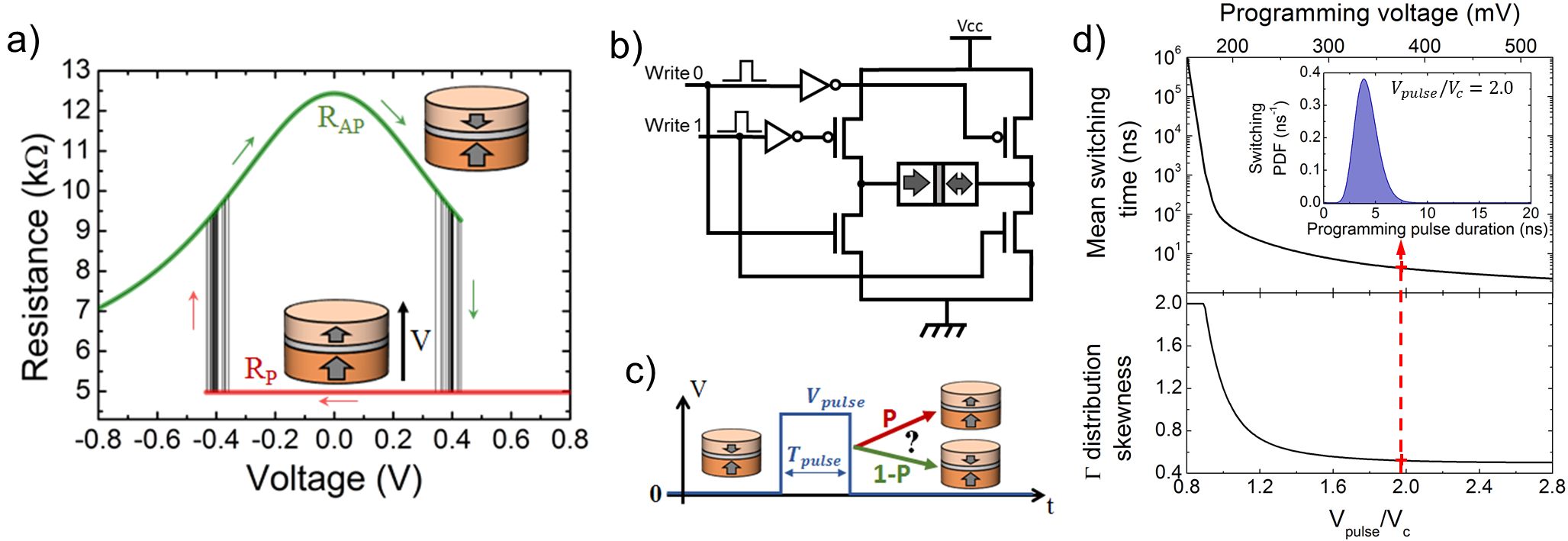}
	\caption{
    (a) Magnetic tunnel junction resistance as a function of voltage, showing stochastic switching phenomenon, from Monte-Carlo simulation. 
    (b) Bidirectional driver circuit. 
    (c) Illustration of the stochastic programming under application of a voltage pulse. 
    (d) Mean switching time and skewness of the gamma probability distribution function for switching, as a function of the programming voltage \Vpulse{} and its ratio to critical voltage $\Vpulse / \Vc$. Inset: Switching PDF for the case $\Vpulse / \Vc = 2.0$.}
	\label{fig:ProgModel}
\end{figure*}

The magnetic tunnel junction (MTJ), the core device of MRAM circuits, is composed of a stack of magnetic and non-magnetic materials, implementing a fixed reference magnet and a bi-stable storage magnet sandwiching a tunnel barrier (Fig.~\ref{fig:ProgModel},a). 
%
%
The resistance of the stack differs for the two relative orientations of the magnets (parallel or anti-parallel), through the tunnel magnetoresistance (TMR) effect. 
The stability of the memory bits is characterized by the energy barrier $\Delta E$ separating the two states. 
Here, $\DeltaE = 70\,\kB T$ is chosen to ensure one Failure In Time rate for a \SI{64}{MB} array. 
%
%
%
%
For this study, we choose values for MTJ parameters reminiscent of a \SI{32}{\nm} node technology based on perpendicular magnetization anisotropy  materials~\cite{khvalkovskiy_basic_2013, chun_scaling_2013}:
diameter of \SI{32}{\nm},
storage layer thickness of \SI{1.3}{\nm},
saturation magnetization of \SI{1.58}{\tesla}/$\mu_{0}$,  
resistance-area product of \SI{4}{\ohm \micro\meter\squared},
\TMR{} of  \SI{150}{\percent}, and 
critical switching voltage of \SI{190}{\mV}.

The state of an MTJ can be switched by passing a current through the stack with appropriate sign and amplitude.
The bit programming can be performed by a bidirectional driver circuit (Fig.~\ref{fig:ProgModel},b) applying a pulse to the device, of given voltage amplitude \Vpulse{} and duration \Tpulse.
One strong particularity of MTJ devices is the stochastic behavior of their programming~\cite{worledge_switching_2010, wang_bit_2012, vincent_analytical_2015, vincent2015spin,ranjan_approximate_2015, sampaio_approximation-aware_2015}.
The junction has a non-100\% probability $P(\Vpulse, \Tpulse)$ to switch (Fig.~\ref{fig:ProgModel},c) when a pulse is applied.
Switching probability distributions can be described by the gamma distribution, whose mean and skewness are derived as a function of the  programming voltage \cite{vincent_analytical_2015}.
The full dependence of the two parameters is presented in Fig.~\ref{fig:ProgModel},d.
For the purpose of this study, complementary MATLAB and SPICE models of the MTJ are developed, that include the full description of the intrinsic stochastic behavior of the device as well as extrinsic variability between devices.
%
%

\section{Interplay between Programming Precision and Energy Cost}
\label{Sec:PrecisionEnergy}

\begin{figure}[ht]
	\centering
	\includegraphics[width=0.9 \linewidth]{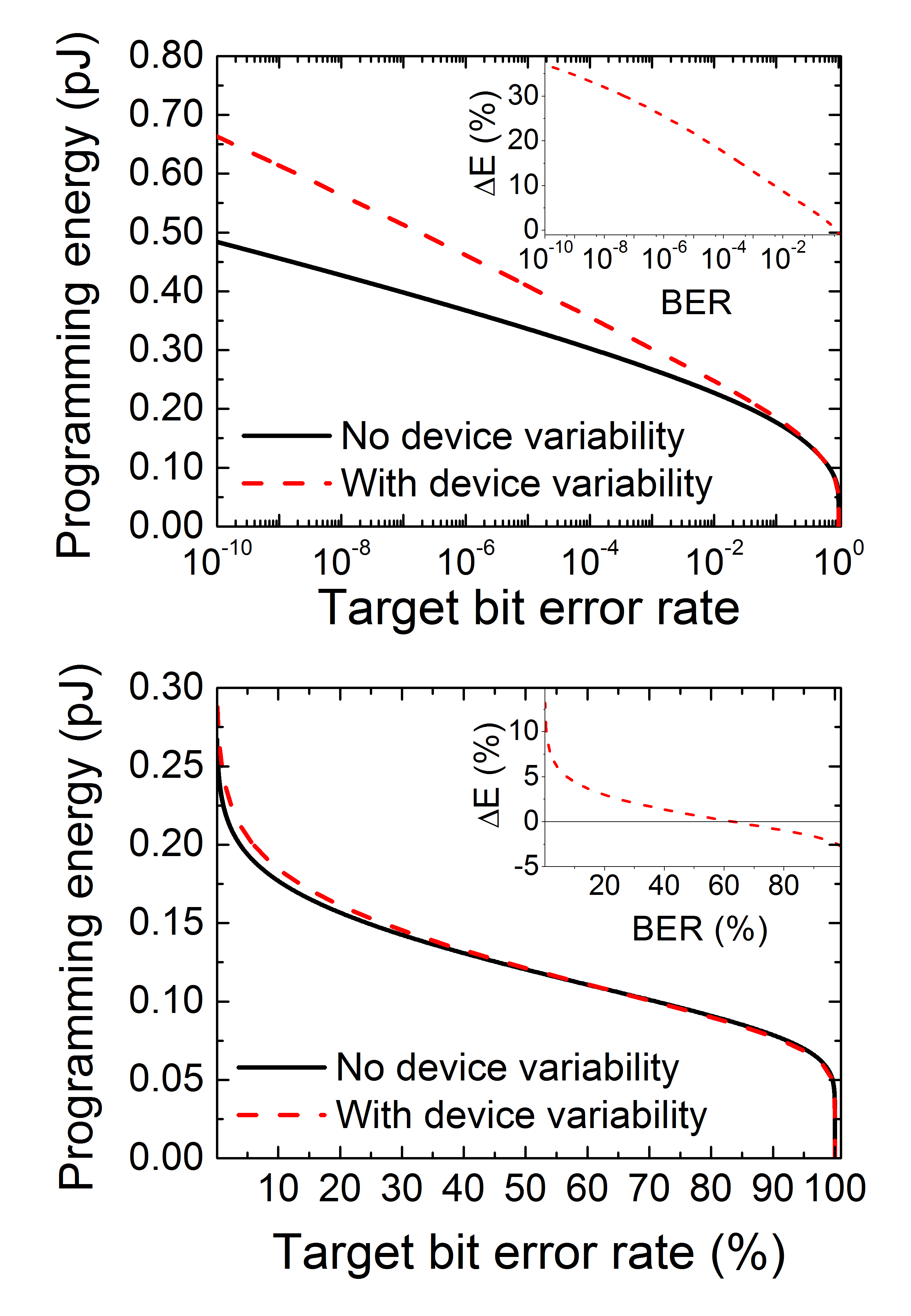}
	\caption{
    Bit programming energy associated with targeted bit error rate, with (red dashed line) or without (black full line) taking into account device variability, in logarithmic (top) and linear (bottom) scales.
    Insets: corresponding relative energy incrase.}
	\label{fig:MCresults}
\end{figure}

In this study, we use the bidirectional driver circuit (Fig.~\ref{fig:ProgModel},b)  to apply pulses with amplitude $\Vpulse = \SI{381}{\mV} = 2.0\,\Vc$. 
Considering no variability in any device of the circuit, we derive from our analytical model that a pulse duration value \Tpulse{} = \SI{15.0}{\ns} is requested to ensure programming with a low bit error rate \BER{} = \num{e-10}.
The associated single bit programming energy is evaluated to \SI{0.48}{\pico\joule}.
Starting from this high-precision reference point, energy savings can be obtained by  increasing  the programming error rate, either by reducing the programming pulse amplitude or its duration.
The strongest energy savings are obtained by a reduction of the pulse duration while keeping the pulse amplitude constant and as high as possible.
Moreover, a pulse duration control strategy can easily be achieved without any modification of the driver circuit or its supply voltage, as it simply requires gating the programming pulse on demand when the application tolerates lower precision storage. 
Fig.~\ref{fig:MCresults} shows the dependence of the programming energy with the target error rate, derived from the model of \cite{vincent_analytical_2015}.
Black solid lines are obtained in the case of a perfect MRAM circuit with no device-to-device variability.
Relaxing the BER from \num{e-10} to \num{e-4} allows for a \SI{37}{\percent} energy reduction.
Best energy savings are expected when considering programming with even lower pulse durations, and higher $\BER \apprge \SI{0.01}{\percent}$.
We then observe a rapid decrease of the programming energy with \BER, as emphasized by Fig.~\ref{fig:MCresults},b.
Compared to programming with \BER~=~\num{e-10}, programming with \BER~=~\num{e-3} allows a \SI{45}{\percent} energy saving, and  \BER{}~=~\SI{1}{\percent} allows a \SI{53}{\percent} energy saving.

Device variability of both CMOS and ST-MRAMs is a crucial issue that can directly increase the global error rate.
To account for these effects, we now simulate the driver circuit using Cadence tools, with the design kit of a \SI{28}{\nm} commercial CMOS technology and our SPICE model of the MTJ devices. 
Process and mismatch variability on CMOS transistors is considered for Monte Carlo simulations, as well as a \SI{5}{\percent} variability on resistance and \TMR{} parameters, typical values of ST-MTJ cells variability~\cite{worledge_switching_2010}.
The consequence of such variability on the programming energy is then evaluated.
The red dashed curve in Fig.~\ref{fig:MCresults} shows the new full dependence of the average programming energy with the target bit error rate.
The relative energy increase is also plotted as insets.
We see that the consequent raise of the programming energy implied by variability is exacerbated when the target bit error rate is decreased. 
Ensuring that the \BER~=~\num{e-10} remains unchanged requests to raise the pulse duration up to \Tpulse~=~\SI{20.5}{\ns}, implying a programming energy increase by \SI{36.8}{\percent}, while only a \SI{17.6}{\percent} increase is necessary for \BER~=~\num{e-4}.
This is due to switching probability becoming increasingly sensitive to variations of the programming parameters when target \BER{} is high.
These results again show the strong cost of high programming precision, and the interest of finding strategies to consider computing with reduced precision whenever possible.




\section{Approximate Programming in a Neural Network}
\label{Sec:NeuralNets}



\begin{figure}[ht]
	\centering
	\includegraphics[width=0.9 \linewidth]{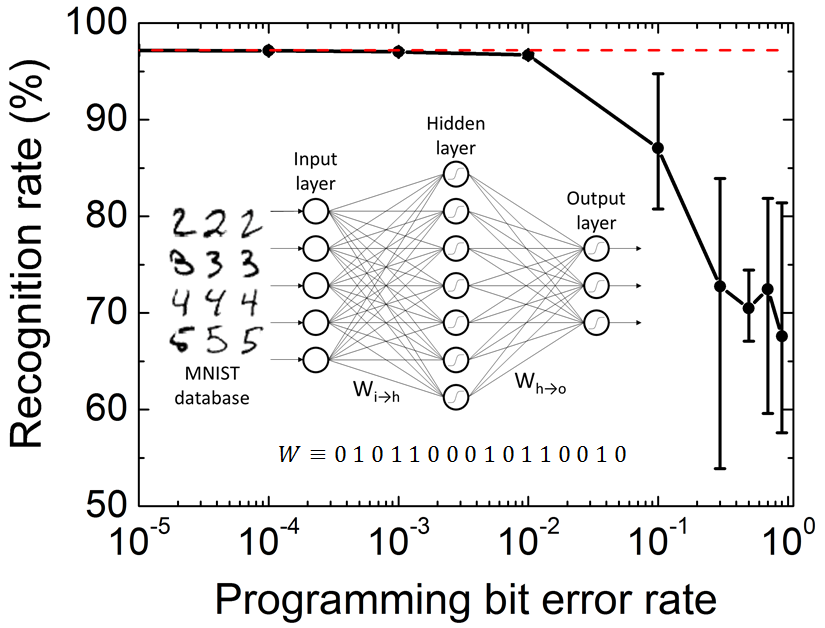}
	\caption{
    Final recognition rate of a 2-layer neural network with 16-bit synaptic weights, as a function of their programming \BER{}, evaluated on the  MNIST dataset.
    Error-bars: dispersion of the results for five different initial conditions.}
	\label{fig:NeuralNet}
\end{figure}

We now consider the use of ST-MRAM in the framework of approximate computing, taking the example of the training of a neural network for an image recognition task.
We consider a simple  two-layer formal neural network with tanh activation function, with \num{784} input neurons, \num{300} hidden neurons, and \num{10} output neurons Fig.~\ref{fig:NeuralNet}).
The network is trained on handwritten digit recognition task using the MNIST dataset,
by updating a  number of synaptic weights along a learning procedure. 
We study the case of weights coded as 16-bit fixed point numbers stored using a ST-MRAM circuit.
For each learning step, a mini-batch of \num{10} images is presented to the ANN, and weight updated following the backpropagation algorithm.
%
%
Learning is performed through five  presentations of the full set of \num{60000} training images.
Five sets of initial weights are used to test the reproducibility and generality of the performance results of the neural network.
For every set of parameters of the approximate memory, we show the obtained average recognition rate, as well as error bars presenting the range of recognition rates obtained over the five tests. 
%
%

In a first phase, we consider programming all the bits of the 16-bit synaptic weights with identical \BER.
Interestingly, in Fig.~\ref{fig:NeuralNet}, we observe that the final recognition rate 
of \SI{97.2}{\percent}
remains unaffected as the \BER{} is increased up to a large \BER~=~\SI{0.1}{\percent}, and is only slightly reduced to \SI{96.7}{\percent} for \BER~=~\SI{1}{\percent}.
Beyond this point, the final recognition rate falls down, and the recognition capabilities become strongly dependent on the initial conditions of the network.
Programming the weights with \BER~=~\SI{0.1}{\percent} or \BER~=~\SI{1}{\percent} represents an energy saving of respectively \SI{54.4}{\percent} and \SI{62.7}{\percent} on the ST-MRAM programming during the learning operation, compared to quasi-deterministic programming (\BER~=~\num{e-10}).

While the energy saving is already substantial, the strategy can be further extended, by choosing different levels of programming precision for each bit, where the memory would prioritize the precision of higher significance bits and progressively increase the BER of lower significance bits. 
In this second phase, we consider a simple two-stage programming strategy by differentiating high significance bits (HSBs), programmed with a \BER{} of \SI{1}{\percent}, and low significance bits (LSBs), programmed with higher \BER. 

\begin{figure}[ht]
	\centering
	\includegraphics[width=0.9 \linewidth]{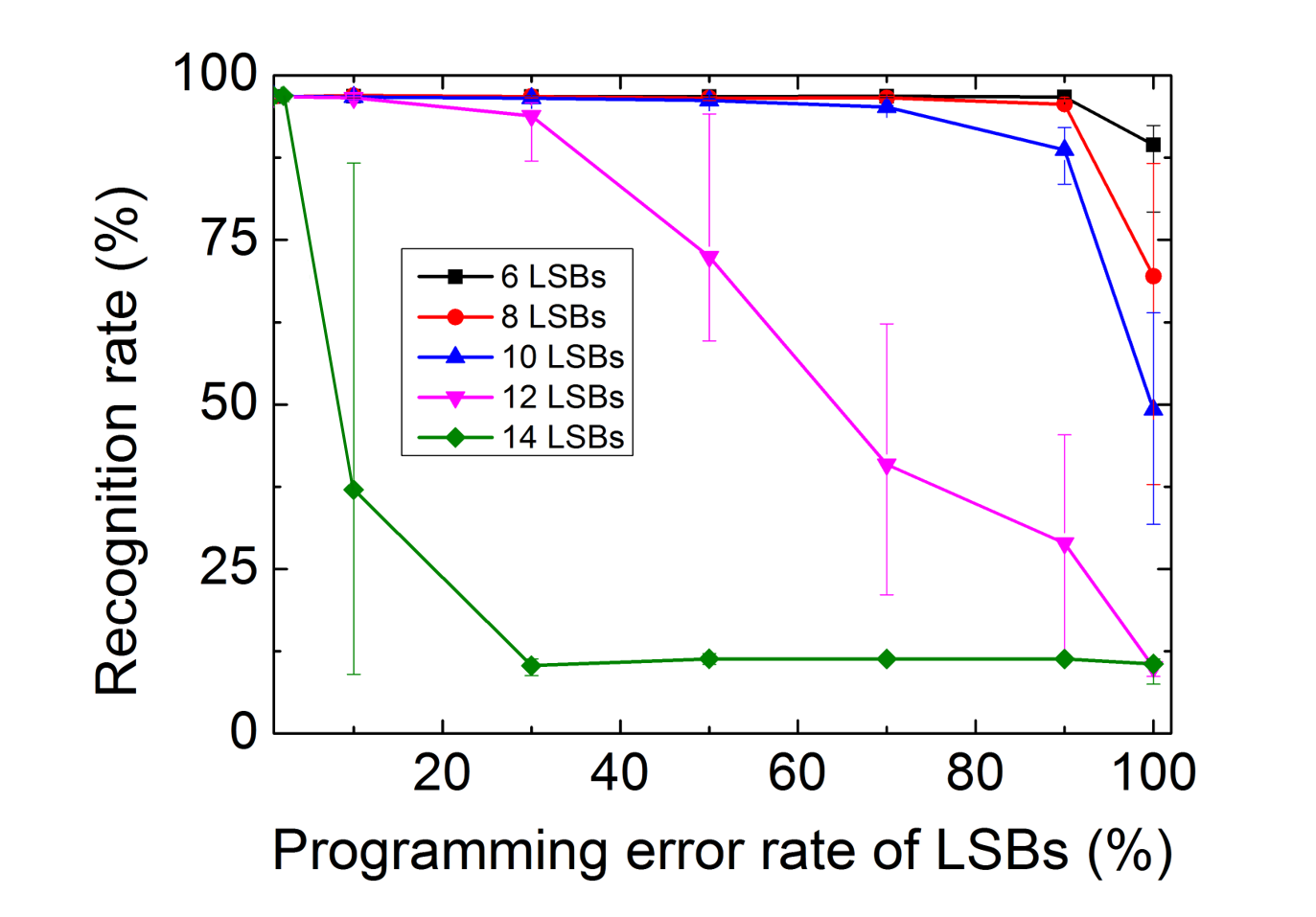}
	\caption{
    Final recognition rate of the 2-layer neural network with 16-bit synaptic weights, as a function of the programming \BER{} of the LSBs, for different numbers of LSB.
    The HSBs are programmed with a \SI{1}{\percent} error rate.}
%
%
	\label{fig:RR_Vs_ER}
\end{figure}

\begin{figure}[ht]
	\centering
	\includegraphics[width=0.9 \linewidth]{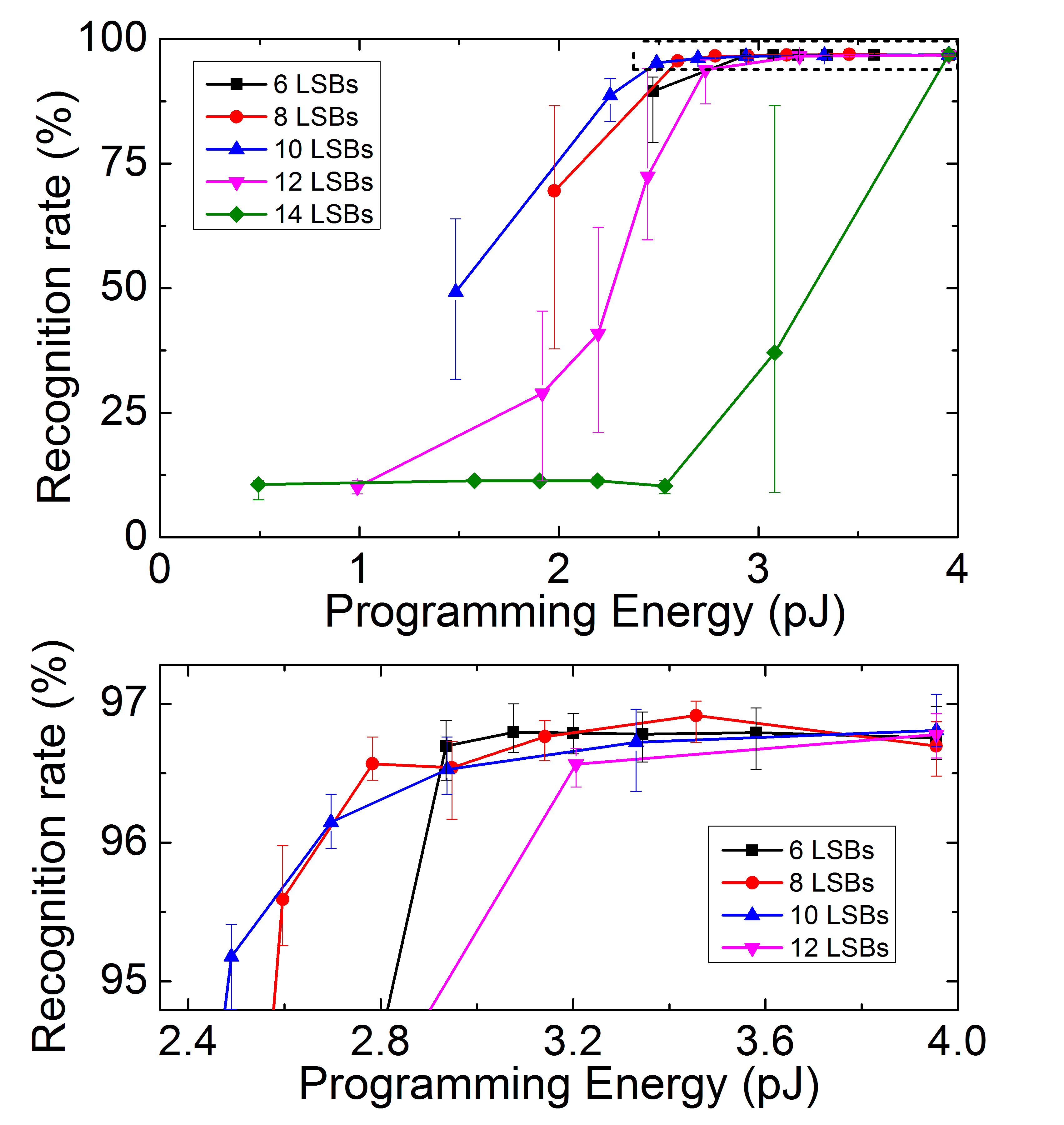}
	\caption{
    (a) Final recognition rate of the 2-layer neural network with 16-bit synaptic weights, as a function of the total programming energy per synaptic weight, for different number of LSBs.
    (b) Detail on the high-performance results in (a).}
%
%
	\label{fig:RR_Vs_Energy}
\end{figure}

The results, presented in Fig.~\ref{fig:RR_Vs_ER}, show that a substantial increase of the \BER{} for LSBs is possible without critical consequences on the final recognition rate of the ANN.
As a reference, results are also given in the case of \BERLSBs~=\SI{100}{\percent}, corresponding to fully random, never updated low significance bits, showing that even a \SI{90}{\percent} \BER{} for the LSBs can substantially increase the performance of the ANN, and discarding the eventuality that LSBs could simply be removed, as can be done in inference-only hardware \cite{hubara2016quantized}.
It is noteworthy that the ANN is resilient to \BERLSBs{} as high as \SI{90}{\percent} for up to \num{8} LSBs, \SI{70}{\percent} for \num{10} LSBs, and \SI{10}{\percent} for \num{12} LSBs.
In the case of \num{14} LSBs, increasing the \BER{} further that \SI{1}{\percent} is immediately detrimental to the performances.  
Again, substantial energy savings can be expected from this strategy.
To visualize the trade-off between programming energy and performances, and find the best compromise between an increase of the number of LSBs and an increase in their \BER, we also transfered the results onto a graph of the ANN final recognition rate versus the total programming energy per synaptic weight, presented on Fig.~\ref{fig:RR_Vs_Energy},a.
%
%
This graph shows that the implementation of the two-stage strategy, is able to bring an extra reduction of the energy expense without affecting very significantly the performances of the ANN.
As emphasized by the high performance window (Fig.~\ref{fig:RR_Vs_Energy},b), best performance-energy ratios are obtained for \num{8} to \num{10} LSBs.
For instance, programming the synaptic weights with \BERHSBs=~\SI{1}{\percent} and \BERLSBs=~\SI{50}{\percent}, while granting a \SI{74.6}{\percent} energy saving compared to quasi-deterministic programming, only reduce the recognition rate of the ANN to \SI{96.2}{\percent}, i.e. less than \SI{1}{\percent} reduction.
Final choice should of course be made depending on the tolerance on the final recognition rate.
%
%

\section{Conclusion}
\label{Sec:CCL}

In this work, we have studied the energy cost of strong programming precision ST-MRAMs, and evaluated the potential energy saving of the programming operation in the framework of disciplined approximate computing.
We highlighted that a control of the programming pulse duration appears as the most efficient handle to tune the programming bit error rate while ensuring the strongest energy reduction. 
Consequently, we proposed a two-fold BER programming strategy that does not require any modification of driver circuits, allowing the application to control which bits should be programmed inaccurately. 
Finally, we showed that using this strategy could result in considerable energy savings when training a hardware neural network.
Taking the simple example of character recognition, \SI{74}{\percent} of the ST-MRAM programing energy can be saved while losing only \SI{1}{\percent} of performance in terms of image recognition.
%
%
%
%
These results confirm that approximate computing techniques can be highly attractive for machine learning applications, and that memory can be a good place where approximate computing is especially useful.

\section*{Acknowledgments}
This work was supported by the  National Research Agency (ANR-10-LABX-0035, ANR-13-JS03-0004-01) and the European Research Council (NANOINFER, ref.: 715872).

\bibliographystyle{myIEEEtran}
\bibliography{Approx_NeuralNets.bib}

\end{document}